\documentclass[12pt]{article}
\usepackage{psfrag}
\usepackage[dvips]{graphicx}
\def\lesssim{\mathrel{\mathpalette\vereq<}}
\def\gtrsim{\mathrel{\mathpalette\vereq>}}
\makeatletter
\def\vereq#1#2{\lower3pt\vbox{\baselineskip1.5pt \lineskip1.5pt
\ialign{$\m@th#1\hfill##\hfil$\crcr#2\crcr\sim\crcr}}}
\makeatother

\begin{document}

\begin{titlepage}

\begin{flushright}
\end{flushright}
\vskip 1.0cm

\begin{center}
{\Large \bf MSW effect for large mixing angles}
\vskip 1.0cm

\def\thefootnote{\fnsymbol{footnote}}
{\large 
Alexander Friedland
}
\vskip 0.5cm

{\it School of Natural Sciences, Institute for Advanced Study, 
Princeton, New Jersey 08540}

\vskip 1.0cm

\abstract{The traditional physical description of neutrino flavor
conversion in the Sun focuses on the notion of resonance.  However,
the resonance picture is valid only in the limit of small mixing
angles $\theta$. For large values of $\theta$, the resonance picture
leads to seemingly paradoxical results. This observation is important
for understanding the physics of neutrino flavor conversion in the
Sun, since the latest solar neutrino data seems to prefer large mixing
angles. Here we review the basic arguments and in particular show that
the resonance does not in general coincide with either the point of
maximal violation of adiabaticity in the nonadiabatic case or the
point of maximal flavor conversion in the adiabatic case. We also
discuss the modified adiabaticity criterion.}

\vskip 2.0cm

\textsl{
  \begin{center}
    Contributed to the Proceedings of the Workshop\\
FRONTIERS IN CONTEMPORARY PHYSICS - II\\
March 2001\\ Vanderbilt University\\
Nashville, TN
  \end{center}
}

\end{center}
\end{titlepage}

\def\thefootnote{\arabic{footnote}}
\setcounter{footnote}{0}


\section{Introduction}

The MSW mechanism \cite{W1977,MS1,MS2} of neutrino flavor oscillations
is a simple and very beautiful idea. The effect arises because of the
coherent forward scattering of the neutrinos on electrons and nucleons
in matter. This scattering makes the ``neutrino index of refraction''
of the media slightly different from one. The remarkable fact is that
this small (and otherwise unobservable) change of the index of
refraction can have a dramatic effect on the oscillation probability
of solar neutrinos. In particular, when the flavor off-diagonal terms
in the neutrino Hamiltonian are small compared to the diagonal terms
in vacuum, the flavor conversion can be almost complete.

It is this possibility of the almost complete conversion that initially
attracted a lot of attention. The phenomenon was understood as a
resonance effect. The flavor conversion was shown to occur at the
depth in the Sun where the density is such that the matter effects
cancel the difference between the diagonal terms in the neutrino
Hamiltonian. 

The recent data \cite{homestake,GALLEX99,SAGE99,SuperK}, however,
together with the latest solar model \cite{BP2000}, indicate that the
preferred value of the mixing angle may be large.  It was recently
emphasized \cite{jumping} that the resonance picture breaks down in
this case. It was shown that the important physical processes, level
jumping or flavor conversion, do not occur at the resonance. The
correct picture is to consider adiabatic or nonadiabatic neutrino
evolution which reduces to the resonance description in the limit
$\theta\ll1$. It is also worth pointing out that this description is
perfectly continuous across the point of the maximal mixing
$\theta=\pi/4$.

In this note we review these developments. The aim is to keep the
presentation as simple as possible, in accordance with the stated
goals of the Workshop to make the Proceedings into a pedagogical
resource. In Section~\ref{standardintro} we present an elementary
introduction to the resonance phenomenon.  
In Section~\ref{sec:questions} we pose several questions that seem to be
difficult to answer within the resonance picture. In
Section~\ref{sec:answers} present the resolutions. 


\section{Neutrino oscillations and the MSW effect}
\label{standardintro}

Let us begin by summarizing the standard treatment of neutrino
oscillations. This introduction is not meant to be exhaustive. In
fact, quite often the best presentations of any idea are written by
people who discovered it or contributed early on, and the case at hand
is no exception, see, {\it e.g.},
\cite{MS1987jetp,Bethe1986,Haxton1986,Parke1986,KP1989review}.

\subsection{Simple vacuum oscillations}
\label{sect:simplevac}

Suppose at some point one creates the electron neutrino
state, for example as a result of a nuclear reaction. Suppose further
that there is a term in the Hamiltonian coupling the electron neutrino
to some other neutrino state. Then, as the neutrino propagates in
vacuum, it will be (partially) converted into that state. Still later,
the state will again be $\nu_e$ and so on, \textit{i. e.}, the flavor
oscillations will take place.

The simplest case is the case of two flavors. The oscillations occur
between the electron neutrino $\nu_e$ and another active neutrino
state\footnote{One should keep in mind 
that the subsequent analysis does not distinguish between
$\nu_\mu$ and $\nu_\tau$. Hence the notation $\nu_\mu$ does not literally mean the
muon neutrino. In fact, the popular oscillation scenario which seems
to be consistent with the atmospheric and solar neutrino data involves
oscillations between $\nu_e$ and a roughly 50--50 mixture of 
 $\nu_\mu$ and $\nu_\tau$.}, conventionally labeled $\nu_\mu$. 
The Schr\"odinger equation for this system is
\begin{equation}
  \label{eq:trivH}
    i \frac{d}{d t}  \left(\begin{array}{c}
      \phi_e\\
      \phi_\mu
   \end{array}\right) =
  \left(\begin{array}{cc}
      H_{ee}&  H_{e\mu}\\
      H_{\mu e} &  H_{\mu\mu}
  \end{array}\right)
\left(\begin{array}{c}
      \phi_e \\
      \phi_\mu
   \end{array}\right).
\end{equation}
The presence of the off-diagonal coupling $H_{e\mu}=H_{\mu e}$ ensures
that the mass eigenstates $|\nu_{1,2}\rangle$ of this system are not
the same as the flavor eigenstates $|\nu_{e,\mu}\rangle$:
\begin{eqnarray}
  \label{eq:massflavor}
  |\nu_1\rangle &=& \cos\theta|\nu_e\rangle - \sin\theta|\nu_\mu\rangle,
 \nonumber\\ 
  |\nu_2\rangle &=& \sin\theta|\nu_e\rangle + \cos\theta|\nu_\mu\rangle.
\end{eqnarray}
Thus, introducing the off-diagonal term in Eq.~(\ref{eq:trivH}) is
equivalent to saying that the mass and flavor eigenstates in the
lepton sector are not aligned. We know this is the case in the quark
section, where the two bases are related by the CKM matrix, hence
it is not unreasonable to expect the misalignment also in the lepton sector.

The coupling $H_{e\mu}$ leads to flavor oscillations as the neutrino
state propagates in space. If at  $x=0$ the state was
purely $\nu_e$, after propagating in vacuum to $x=L$ it will be
measured as $\nu_\mu$ with the probability
\begin{equation}
  \label{eq:simplePvac}
  P(\nu_e\rightarrow\mu_\mu) = \sin^2 2\theta \sin^2\left(\frac{\Delta}{2}L\right).
\end{equation}
Here $\Delta$ is one half of the mass splitting,
$\Delta\equiv\sqrt{H_{e\mu}^2+(H_{\mu\mu}-H_{ee})^2/4}=\Delta m^2/2E_\nu$. 
Clearly, $\Delta\sin 2\theta=H_{e\mu}$,
$\Delta\cos2\theta=(H_{\mu\mu}-H_{e e})/2$.  
Since all oscillation phenomena depend only on 
$(H_{\mu\mu}-H_{ee})/2$ and not on $H_{\mu\mu}$ or  
$H_{e e}$ separately, we shall often drop the constant term in the
Hamiltonian
\begin{eqnarray}
  \label{eq:dropc}
  H &=&
  \frac{H_{\mu\mu}+H_{ee}}{2}+\left(\begin{array}{cc}
      -(H_{\mu\mu}-H_{ee})/2 &  H_{e\mu}\\
      H_{\mu e} &  (H_{\mu\mu}-H_{ee})/2
    \end{array}\right)\nonumber\\
  &\longrightarrow&
  \left(\begin{array}{cc}
      -(H_{\mu\mu}-H_{ee})/2 &  H_{e\mu}\\
      H_{\mu e} &  (H_{\mu\mu}-H_{ee})/2
    \end{array}\right)=
  \left(\begin{array}{rr}
      -\Delta\cos 2\theta &  \Delta\sin 2\theta\\
      \Delta\sin 2\theta &  \Delta\cos 2\theta
    \end{array}\right).
\end{eqnarray}  

\subsection{The resonance and the MSW effect}
\label{sect:simpleres}

\subsubsection{Matter effects}

The above picture describes the evolution of the neutrino state in
vacuum. One might be tempted to also use the same formulas for the
evolution in matter. Indeed, it might seem that the presence of matter
has no effect on the neutrino evolution. After all, the scattering
cross section for the electron neutrino $\sigma\sim G_F^2 E_{\rm
c.m.}^2$ is so small for solar neutrinos ($E_{\rm lab}\sim (0.2-15)$
MeV), that the vast majority of them go right through the Sun undeflected.

Nevertheless, the presence of matter does affect the neutrino evolution in an
important way. Since the medium is transparent to neutrinos, it is
helpful to think in the language of ``neutrino optics''. The presence
of the medium can lead to a index of refraction for the
neutrinos. Since the index of refraction is a phase phenomenon -- it is
related to the \emph{amplitude} of the coherent forward scattering --
the effect involves the first power of $G_F$. Wolfenstein pointed out
\cite{W1977} that this coherent forward scattering leads to the
addition of the extra terms to the diagonal elements of the
Hamiltonian, $\sqrt{2} G_F (N_e - N_n/2)$ for the electron neutrinos
and $\sqrt{2} G_F ( - N_n/2)$ for the other active species. Thus,
\begin{equation}
  \label{eq:Hmat}
  H_{\rm mat} = {\rm const} +\left(\begin{array}{cc}
      A-\Delta\cos 2\theta &  \Delta\sin 2\theta\\
       \Delta\sin 2\theta &  \Delta\cos 2\theta-A
  \end{array}\right)
\end{equation}
where $A\equiv G_F N_e/\sqrt{2}$. The effect of the neutrino
interaction with matter is to modify the mixing angle and the mass
splitting in matter,
\begin{equation}
  \label{eq:Hmat2}
  H_{\rm mat} = {\rm const} +\left(\begin{array}{cc}
      \Delta_m\cos 2\theta_m &  \Delta_m\sin 2\theta_m\\
       \Delta_m\sin 2\theta_m &  \Delta_m\cos 2\theta_m
  \end{array}\right).
\end{equation}
where
\begin{eqnarray}
  \label{eq:relateDelta_m}
\Delta_m &=&
  \sqrt{A^2-2 A\Delta \cos 2\theta+\Delta^2},\\
  \label{eq:relatetan_m}
  \tan 2\theta_m &=& \frac{\Delta\sin 2\theta}{\Delta\cos 2\theta-A}.
\end{eqnarray}

To find the oscillation probability in a medium of constant density, one can use
Eq.~(\ref{eq:simplePvac}) with $\Delta\rightarrow\Delta_m$,
$\theta\rightarrow\theta_m$. This results, for example, in the suppression of
the oscillations in the solar core for $\Delta\lesssim 10^{-5}$ eV$^2$/MeV
\cite{W1977}.

\subsubsection{Resonance: the idea}

The true importance of the matter effects for addressing the solar
neutrino problem was not realized until 1985, when the so-called
resonance phenomenon was discovered \cite{MS1,MS2}. It was known at
the time that the Chlorine and Gallium data showed a 50-70\%
suppression of the solar neutrino flux. One way to get this
suppression was to have a large mixing angle and oscillations in
vacuum. The resonance idea provided another way. It showed that a
large flavor conversion could occur even if in vacuum $ |H_{e\mu}|\ll
|H_{\mu\mu}-H_{e e}|$.

\begin{figure}[t]
  \begin{center}
   \includegraphics[angle=0, width=0.6\textheight]{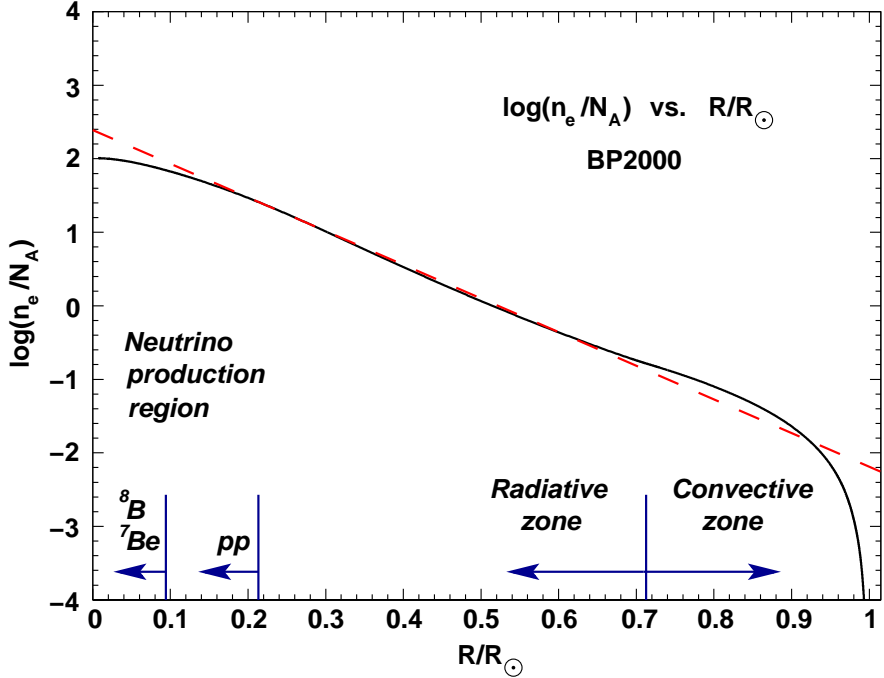}
    \caption{Electron density in the Sun according to the BP2000
      \cite{BP2000} solar model. The dashed line shows the best-fit
      exponential $n_e/N_A=245\exp(-10.54R/R_\odot)$.}
    \label{fig:profile}
  \end{center}
\end{figure}

In a nutshell, the idea is to have the matter term cancel the
difference of the diagonal elements at some depth in the Sun.  If, at
some depth $\tilde x$, $|A(\tilde x) - \Delta \cos2\theta|<\Delta\sin
2\theta$, the conversion will take place according to
$id\phi_\mu/dt\simeq\Delta\sin 2\theta\phi_e$. Notice that the
condition 
\begin{equation}
  \label{eq:resonance}
  A(x_0) = \Delta \cos 2\theta
\end{equation}
{\it a priori} does not require any large fine-tuning of the neutrino
parameters, since the electron density in the Sun smoothly varies over
several orders of magnitude (see Figure~\ref{fig:profile}).

To get enough conversion, the distance $\delta x$ over
which the cancellation persists must be long enough so that a significant
fraction of the oscillation cycle can be completed,
\begin{eqnarray}
  \label{eq:simple_res}
  \Delta\sin 2\theta (\delta x) \gtrsim 1.
\end{eqnarray}
If one assumes that the profile around the resonance $x=x_0$ is
approximately linear $A(x)\simeq A(x_0)+A'(x_0)\delta x$,
the width of the resonant region can be written as
\begin{equation}
  \label{eq:width}
  2\delta x \simeq \frac{2\Delta\sin 2\theta}{A'(x_0)}
\end{equation}
and Eq.~(\ref{eq:simple_res}) becomes
\begin{eqnarray}
  \label{eq:simple_res2}
  \frac{2\Delta^2\sin^2 2\theta}{A'(x_0)} \gtrsim 1.
\end{eqnarray}

Whenever the condition in Eq.~(\ref{eq:simple_res2}) is satisfied, one can
expect the conversion to be large.

\subsubsection{Resonance: level crossing}
\label{sect:levelcrossing}

\begin{figure}[t]
  \begin{center}
    \psfrag{R/Rsun}{\large $R/R_\odot$}
    \psfrag{nue}{\Large $\sim\nu_e$}
    \psfrag{numu}{\Large $\sim\nu_\mu$}
    \psfrag{0}{\large 0}
    \psfrag{1}{\large 1}
    \psfrag{0.2}{\large 0.2}
    \psfrag{0.4}{\large 0.4}
    \psfrag{0.6}{\large 0.6}
    \psfrag{0.8}{\large 0.8}
    \includegraphics[angle=0, width=0.7\textwidth]{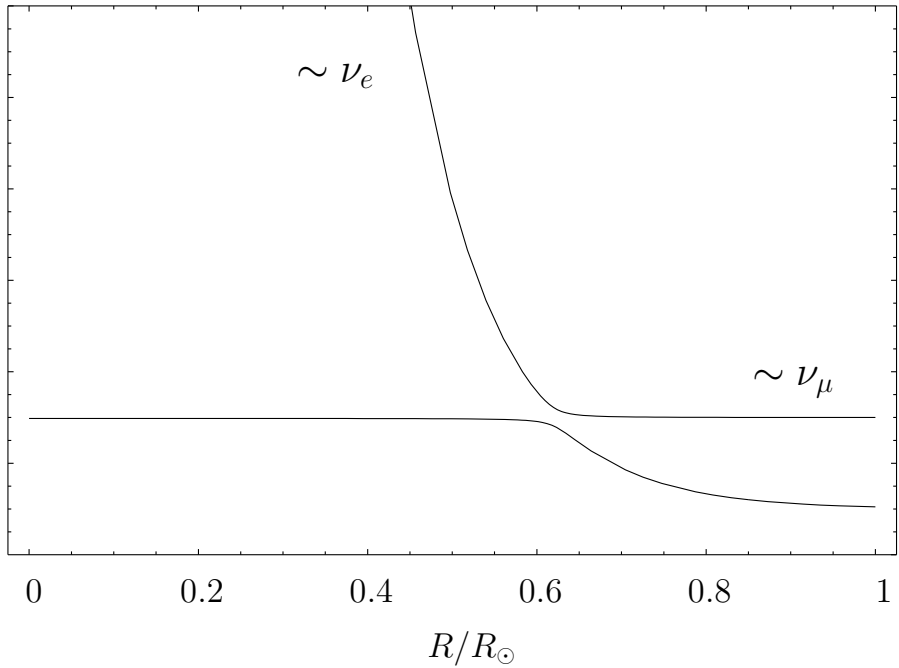}
    \caption{The level crossing in the Sun (small mixing angle).}
    \label{fig:crossing}
  \end{center}
\end{figure}

On closer inspection, it turns out that the conversion is not only
large, but can be almost complete. This can most easily understood as
the case of adiabatic level crossing, a phenomenon well known in
quantum mechanics. If parameters in the Hamiltonian vary slowly
enough, the state that was initially created as the eigenstate of the
instantaneous Hamiltonian (henceforth, ``the matter mass eigenstate'')
will follow the changing eigenstate. In the case of solar neutrinos
with a small mixing angle and $\Delta\leq 10^{-5}$ eV$^2$/MeV, the
flavor eigenstate $\nu_e$ in the core almost coincides with the heavy
matter mass eigenstate $\nu_2$. The heavy mass eigenstate is, in turn,
smoothly ``connected'' to the heavy eigenstate in vacuum which is
predominantly composed of $\nu_\mu$. The situation is illustrated in
Fig. \ref{fig:crossing}.

To see explicitly how the adiabatic evolution works, let us
transform the evolution equation, Eq.~(\ref{eq:trivH}), from the
flavor basis $(\nu_e,\nu_\mu)$ to the basis of the matter mass
eigenstates $(\nu_1,\nu_2)$,
$$\phi=V^\dagger\psi.$$
 Remembering that
the rotation matrix 
\begin{eqnarray}
  \label{eq:V}
  V=
  \left(\begin{array}{rr}
    \cos\theta_m & -\sin\theta_m \\
    \sin\theta_m & \cos\theta_m 
  \end{array}\right)
\end{eqnarray}
is position dependent, we get
\begin{eqnarray}
  \label{eq:instantH}
  i \frac{d}{d x} (V^\dagger \psi) &=& H V^\dagger
  \psi, \nonumber\\
  i \frac{d\psi}{d x}  &=& V H V^\dagger \psi - 
i V \frac{d V^\dagger}{d x} \psi.
\end{eqnarray}
Since $V H V^\dagger=\mbox{diag}(-\Delta_m,+\Delta_m)$ and from
Eq.~(\ref{eq:V}) 
\begin{displaymath}
  V\frac{d V^\dagger}{d x}=
  \left(\begin{array}{rr}
    0 & 1 \\
    -1 & 0
  \end{array}\right)\frac{d \theta_m}{d x},
\end{displaymath}
we obtain the desired evolution equation in the basis of the matter
mass eigenstates \cite{MS1987jetp},
\begin{eqnarray}
  \label{eq:massbasis}
  \frac{d}{d x}  
  \left(\begin{array}{c}
      \psi_1 \\
      \psi_2
   \end{array}\right) =   
  \left(\begin{array}{cc}
      i \Delta_m & -d\theta_m/d x \\
      d\theta_m/d x & -i \Delta_m
    \end{array}\right) 
  \left(\begin{array}{c}
      \psi_1 \\
      \psi_2
   \end{array}\right).
\end{eqnarray}

This shows that the coupling between the matter mass eigenstates depends on
$d\theta_m/d x$ and, hence, vanishes when the density change along the neutrino
trajectory is very gradual. More precisely, the relevant quantity is the ratio
$\Delta_m/(d\theta_m/d x)$, which at the resonance point equals $2\Delta^2\sin^2
2\theta/A'(x_0)$. Thus the condition for adiabaticity $\Delta_m/(d\theta_m/d x)
\gtrsim 1$ is precisely the same as Eq.~(\ref{eq:simple_res2}), as should be
expected since they both describe the same physics.

\subsubsection{Vacuum oscillation solutions}

If matter effects are important in the Sun, what about the vacuum oscillation
solutions to the solar neutrino problem? Wouldn't the matter effects also modify
the neutrino evolution in that case?

This question can be answered in several steps. First, the vacuum
oscillation solutions are characterized by the values of $\Delta m^2$ in the
range $10^{-11}-10^{-9}$ eV$^2$. For $\Delta m^2$ in this range, the matter
effects are obviously important in the core: the term $A(x)$ completely
dominates in Eq. (\ref{eq:Hmat}), leading to the complete suppression of the
oscillations there.

Next, consider what happens in the outer layers on the Sun. If $\Delta/E_\nu
\sim 10^{-11}$ eV$^2$/MeV, the off-diagonal term in the Hamiltonian is so small
that the neutrino does not have any time to oscillate into $\nu_\mu$ while still
inside the Sun. In other words, the condition (\ref{eq:simple_res2}) in this
case is badly violated and the neutrino evolution is \emph{extremely
nonadiabatic} (flavor conserving). In the language of the matter mass eigenstates,
this corresponds to the probability of level jumping $P_c=\cos^2\theta$. One
then expects the vacuum oscillations to start up near the surface of the
Sun\footnote{A detailed analysis of the oscillation phase was recently done in
\cite{LisiPetcov}. The value of the oscillation phase in the vacuum oscillation
region was found to be as if the vacuum oscillations started up at $r=0.87
R_\odot$.}.

Finally, it is worth mentioning that this conclusion does not extend to
$\Delta/E_\nu \sim 10^{-9}$ eV$^2$/MeV. A careful analysis shows that the
neutrino evolution in this range is not extremely nonadiabatic, unlike
previously thought, {\it i. e.}, $P_c<\cos^2\theta$. For details, see
\cite{vacuummsw}.

\section{Questions}
\label{sec:questions}

Above we have described several key features of the neutrino
evolution in the Sun. They form the well established physical picture
of the MSW mechanism as the ``resonant amplification
phenomenon''. This picture has become the standard lore over the years.
It can be summarized as follows:
\begin{itemize}
\item If the evolution is adiabatic, the flavor content of the
  neutrino state changes as the neutrino traverses the Sun. The flavor
  conversion occurs around the resonance layer, see Eq.~(\ref{eq:resonance}).
\item To check whether the evolution is adiabatic or not, one should
  evaluate the adiabaticity condition. This condition becomes the most
  critical at the resonance, Eq.~(\ref{eq:simple_res2}). If the
  evolution is nonadiabatic, the jumping between the matter mass
  eigenstates takes place. This jumping also happens mostly in the resonance
  layer.
\end{itemize}

This picture is remarkably simple and at the same time very successful
in describing the neutrino evolution when the mixing angle is
small. It, however, encounters serious difficulties explaining physics
in other cases. We shall next pose several questions that expose the
limitations of this picture.

\subsection{The relevant part of the density profile}

In the preceding discussion we assumed that the density profile was
linear. It may not be immediately obvious why this assumption is
justified. After all, the electron density throughout the Sun,
generally speaking, is not linear. As Figure~\ref{fig:profile} shows,
it falls off more or less exponentially in the radiative zone, while
deviating from the exponential near the solar edge and also in the
core.

More precisely, we assumed that the density can be well approximated
by a linear function in some neighborhood $(x_0-\delta x,x_0+\delta
x)$ of the resonance point,  where 
$\delta x=2\Delta\sin 2\theta/A'(x_0)$ is the previously introduced width of
the resonance region. Obviously, this works well when
 $A''(x_0) (\delta x)^2/2 \ll A'(x_0) (\delta x)$, or
$$\Delta\sin 2\theta\ll (A'(x_0))^2/A''(x_0).$$

We can see that, when the mixing angle $\theta$ is small, the width of
the resonance region is also small, so that the linear approximation
for the density should be good. The important point is that all the
``interesting'' physics, {\it i. e.} the flavor conversion or level
jumping, happens within this region. The jumping probability in this
case can be found using the analytical expression for the linear
density profile,
\begin{eqnarray}
  \label{eq:linearPc}
  (P_c)_{\rm (lin)}=
  \exp\left(-\pi \frac{\Delta^2 \sin^2 2\theta}{A'|_{A=\Delta}}\right).
\end{eqnarray}
(Notice that the expression in the exponent is up to a factor just the
adiabaticity parameter.)
The details of the profile outside this region are not so important,
so long as the density there is sufficiently different from resonant.

The question is: what happens when the mixing angle is large?
According to Eq~(\ref{eq:width}), the resonance region in that case is
wide and so the linear approximation may no longer apply. While the
analytical expression for the infinite exponential profile
\cite{petcovanalyt,Toshev},
\begin{equation}
  \label{eq:expPc}
  P_c=\frac{e^{4\pi r_0\Delta \cos^2\theta}-1}{e^{4\pi r_0\Delta}-1},
\end{equation}
is not limited to the case of small $\theta$, it is not clear \emph{a
priori} if it can be applied for solar neutrinos, since there we
are not dealing with an infinite exponential. If the ``interesting''
physics, in the sense defined before, happens on the part of the
trajectory which does not lie within the exponential part of the
profile, we have no justification to use Eq.~(\ref{eq:expPc}).

The question is not just academic, but has a direct application for
the present-day solar neutrino analysis. Recent solar neutrino fits
indicate that the so-called Large Mixing Angle (LMA)
solution, characterized by $\Delta m^2\sim 10^{-5}-10^{-4}$ eV$^2$ and 
$\tan^2\theta\sim 0.1-0.9$, gives a marginally better fit than the
small mixing angle (SMA) solution. As the mixing angle approaches $\pi/4$, the
resonance defined in Eq.~(\ref{eq:resonance}) occurs closer and closer
to the solar surface. Does this mean we need to know the details of
the profile in the convective zone of the Sun to reliably compute the
neutrino survival probability for the LMA solution? The resonance
amplification picture says ``yes'', while the numerical computations
say ``no''.

\subsection{The case of ``inverted hierarchy''}

The next issue is understanding the evolution in the case $\Delta<0$
(``inverted hierarchy''). This means the heavy mass eigenstate in
vacuum is made up predominantly of $\nu_e$. Since in this case no
resonance occurs, it received comparatively little attention in the
literature until recently. At the same time, this case raises several
interesting questions that are not easy to answer within the traditional
resonance framework described above.

First of all, notice that the mixing angles at the production point in
the core and in vacuum are different, just like in the $\Delta>0$
case. This means that if the evolution is adiabatic, the flavor
composition of the neutrino state changes as it travels through the
Sun. Where does this change occur if there is no resonance?

Next, consider the limit of small $\Delta$. For $\Delta\sim
10^{-11}-10^{-10}$ eV$^2$/MeV one expects vacuum oscillations which
are no different then in the $\Delta>0$ case. Namely, the flavor
oscillations in the solar core are suppressed and in the region close
to the solar surface the off-diagonal terms in the Hamiltonian
(\ref{eq:Hmat}) are too small to allow a significant flavor conversion
within the Sun. Hence, viewed in the mass basis, the evolution in this
case is extremely nonadiabatic. The jumping between the matter mass
eigenstates does occur, even if there is no resonance. How can this be?

\section{Matter effects and large mixing angles}
\label{sec:answers}

\subsection{Connecting $\Delta<0$ and $\Delta>0$ parts of the parameter
  space}

The two cases, $\Delta<0$ and $\Delta>0$, the way they were described
up to this point, might appear qualitatively very
different. Nevertheless, it can be easily demonstrated that they are
continuously connected.

To see this, notice that the situations 1) $\Delta<0$,
$\theta\leq\pi/4$ and 2) $\Delta>0$, $\theta\geq\pi/4$ are physically
the same \cite{FLM1996,ourdarkside}. This can be easily seen from
Eqs.~(\ref{eq:trivH}) and (\ref{eq:dropc}). The case
$\theta\rightarrow\pi/2-\theta$ reduces to $\Delta\rightarrow -\Delta$
by redefining the sign of, say, $\phi_\mu$.
Hence the physical parameter space can be continuously parameterized by
keeping the sign of $\Delta$ fixed and varying $\theta$ from 0 to
$\pi/2$. This means that there should be a unifying physical
description of the MSW effect that \emph{continuously} incorporates
the resonant $(\theta<\pi/4)$ and the nonresonant $(\theta>\pi/4)$
cases.

\subsection{The points of maximal violation of adiabaticity and
  maximal flavor conversion do not coincide with the resonance}

Such description is not difficult to construct. The two questions we
must be able to answer are:
\begin{itemize}
\item In the adiabatic case, what defines the center of the region
  where the flavor conversion takes place?
\item In the nonadiabatic case, what defines the center of the region
  where the jumping between the mass eigenstates occurs?
\end{itemize}
Once again, the description we are seeking should be continuous
across the maximal mixing ($\theta=\pi/4$) and should reduce to the
standard resonance picture in the small $\theta$ limit.

The answers to both questions turn out to be straightforward. First,
consider the adiabatic case. In this regime the neutrino state is
``glued'' to the changing mass eigenstate. Hence, one can define the
point a which the mass basis rotates at the maximal rate with respect
to the flavor basis to be the point where the flavor composition of
the state changes at the fastest rate. In other words, we need the
point where $\dot\theta_m\equiv d\theta_m/dx$ is maximal.

\begin{figure}[t]
  \begin{center}
    \psfrag{psi}{{\large $|\psi_2|^2$}}
    \psfrag{r, 10^6 km}{{\large $x$, $\times 10^6$ km}}
    \psfrag{pi/3}{{\large $\theta=\pi/3$}}
    \psfrag{pi/4}{{\large $\theta=\pi/4$}}
    \psfrag{pi/6}{{\large $\theta=\pi/6$}}
    \psfrag{pi/60}{{\large $\theta=\pi/60$}}
    \psfrag{1}{{\large 1}}
    \psfrag{1.4}{{\large 1.4}}
    \psfrag{1.2}{{\large 1.2}}
    \psfrag{1.0}{{\large 1.0}}
    \psfrag{0.8}{{\large 0.8}}
    \psfrag{0.6}{{\large 0.6}}
    \psfrag{0.4}{{\large 0.4}}
    \psfrag{0.2}{{\large 0.2}}
    \includegraphics[angle=0, width=.7\textwidth]{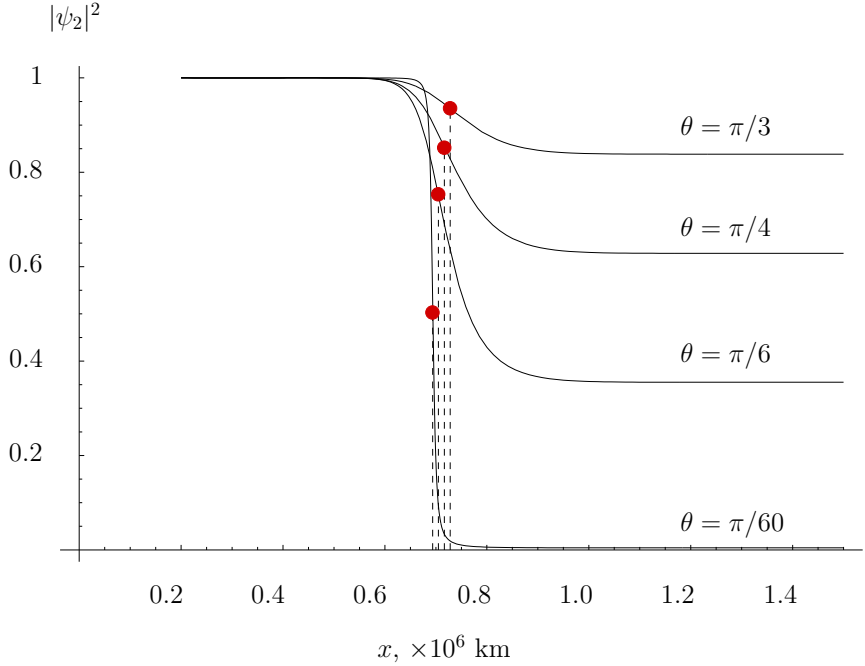}
    \caption{Neutrino state evolution in the case of the 
      infinite exponential density profile for $\Delta m^2/E_\nu=10^{-9}$
      eV$^2$/MeV. The plot shows the probability of finding the
      neutrino in the heavy matter mass eigenstate $\nu_2$ as a function of
      position $x$, for four different values of the vacuum mixing
      angle. The points of the maximal violation of adiabaticity,
      as predicted by Eq. (\ref{eq:minexp2}), are marked.}
    \label{fig:jumpingregion_exp}
  \end{center}
\end{figure}

To get some understanding of where this point lies, let us find it
explicitly for the infinite exponential profile
$A(x)\propto\exp(-x/r_0)$. Using  $A'(x)=-A(x)/r_0$ and
\begin{eqnarray}
  \label{eq:thetadot}
  \dot\theta_m&=&\frac{\sin^2 2\theta_m}{2\Delta \sin 2\theta}
  \frac{d A}{d x},\\
  A&=&\frac{\Delta \sin(2\theta_m-2\theta)}{\sin 2\theta_m},
\end{eqnarray}
we find that 
$\dot\theta_m\propto\sin2\theta_m\sin(2\theta_m-2\theta)=(\cos2\theta-\cos(4\theta_m-2\theta))/2$,
and so the maximum occurs when
\begin{equation}
  \label{eq:maxangle}
  \theta_m^{\rm max}=\pi/4+\theta/2,
\end{equation}
{\it i. e.}, when $\theta_m$ is halfway between its value deep in the core
$(\pi/2)$ and its value in vacuum $(\theta)$. The corresponding
value of the matter term at this point is
\begin{equation}
  \label{eq:maxconversion}
  A(\theta_m^{\rm max})=\Delta.
\end{equation}

Notice that conditions in Eqs.~(\ref{eq:maxangle}) and
(\ref{eq:maxconversion}) are completely general, valid for \emph{all}
values of the mixing angle. Notice also that they agree with the
resonance description
\begin{eqnarray}
  \label{eq:rescond}
  A(x_0)=\Delta\cos 2\theta,\\
  \theta_m(x_0)=\pi/4,
\end{eqnarray}
in the limit $\theta\ll 1$, as expected. Thus, the resonance
amplification interpretation is just a particular limit of this
general physical picture.

Now consider the nonadiabatic case. The ``level jumping'' region will
be centered around the point of maximal violation of adiabaticity,
which corresponds to the minimum of $\Delta_m/\dot\theta_m$, as
discussed in Sect. \ref{sect:levelcrossing}. Since
$\dot\theta_m\rightarrow\infty$ deep inside the Sun and also in
vacuum, the minimum of $\Delta_m/\dot\theta_m$, unlike the resonance,
exists for \emph{all} values of $\theta$.

Once again, let us demonstrate this on the example of the exponential
profile. A short calculation shows that the minimum in this case
occurs when
\begin{eqnarray}
  \label{eq:minexp1}
  \cot(2\theta_m-2\theta)+2\cot(2\theta_m)=0,
\end{eqnarray}
or 
\begin{eqnarray}
  \label{eq:minexp2}
  A=\Delta \frac{\cos 2\theta+\sqrt{8+\cos^2 2\theta}}{4}.
\end{eqnarray}
Thus, the density at the center of the nonadiabatic part of the
neutrino trajectory varies between $A=\Delta$ ($\theta\rightarrow0$)
and  $A=\Delta/2$ ($\theta\rightarrow\pi/2$)\footnote{This is simple
  to understand. The term $1/\dot\theta_m$ has a minimum at $A=\Delta$,
  and since $\Delta_m$ is falling with $x$, the minimum of the product
  in the case of large $\theta$ is shifted towards somewhat smaller
  $A$. When $\theta$ is small, the minimum of $1/\dot\theta_m$ is very
  sharp and completely determines the minimum of the product.}. 
The resonance description again only works when $\theta\ll 1$. As
$\theta\rightarrow\pi/4$, the resonance moves to infinity, while the
true point of maximal violation of adiabaticity moves only to the
point where $A=\Delta/\sqrt{2}$.

The situation is illustrated in Fig. \ref{fig:jumpingregion_exp},
which shows the probability of finding the neutrino in the heavy mass
state $\nu_2$ as a function of the distance $x$. The parameters of the
exponential were taken from the fit line in Fig. \ref{fig:profile} and
$\Delta m^2/E_\nu=10^{-9}$ eV$^2$/MeV. Three large values of the
mixing angle ($\theta=\pi/6,\pi/4,\mbox{ and } \pi/3$) and one small
value ($\theta=\pi/60$) were chosen. The dashed lines and dots mark
the points where adiabaticity is maximally violated, as predicted by
Eq. (\ref{eq:minexp2}). One can see that the partial jumping into the
light mass eigenstate in all four cases indeed occurs around the marked
points.

\begin{figure}[p]
  \begin{center}
    \psfrag{tan}{{\large $\tan^2 \theta$}}
    \psfrag{dm2}{{\large $\Delta m^2/E_\nu$, (eV$^2$/MeV)}}
    \psfrag{0}{{\large $1$}}
    \psfrag{1}{{\large $10$}}
    \psfrag{2}{{\large $10^2$}}
    \psfrag{3}{{\large $10^3$}}
    \psfrag{-1}{{\large $10^{-1}$}}
    \psfrag{-2}{{\large $10^{-2}$}}
    \psfrag{-3}{{\large $10^{-3}$}}
    \psfrag{7}{{\large $10^{-7}$}}
    \psfrag{8}{{\large $10^{-8}$}}
    \psfrag{9}{{\large $10^{-9}$}}
    \psfrag{10}{{\large $10^{-10}$}}
    \psfrag{11}{{\large $10^{-11}$}}
    \psfrag{Pc=0.9}{{\large $P_c=0.9$}}
    \psfrag{0.7}{{\large $0.7$}}
    \psfrag{0.5}{{\large $0.5$}}
    \psfrag{0.3}{{\large $0.3$}}
    \psfrag{0.1}{{\large $0.1$}}
    \psfrag{0.01}{{\large $10^{-2}$}}
    \psfrag{0.001}{{\large $10^{-3}$}}
    \psfrag{Adiabatic}{\textsf{\Large Adiabatic}}
    \psfrag{Nonadiabatic}{\textsf{\Large Nonadiabatic}}
    \psfrag{Eq1}{{\large Eq. (\ref{eq:oldad0})}}
    \psfrag{Eq2}{{\large Eq. (\ref{eq:oldad})}}
    \psfrag{Eq3}{{\large Eq. (\ref{eq:adiabaticity})}}
    \includegraphics[angle=0,width=.9\textwidth]{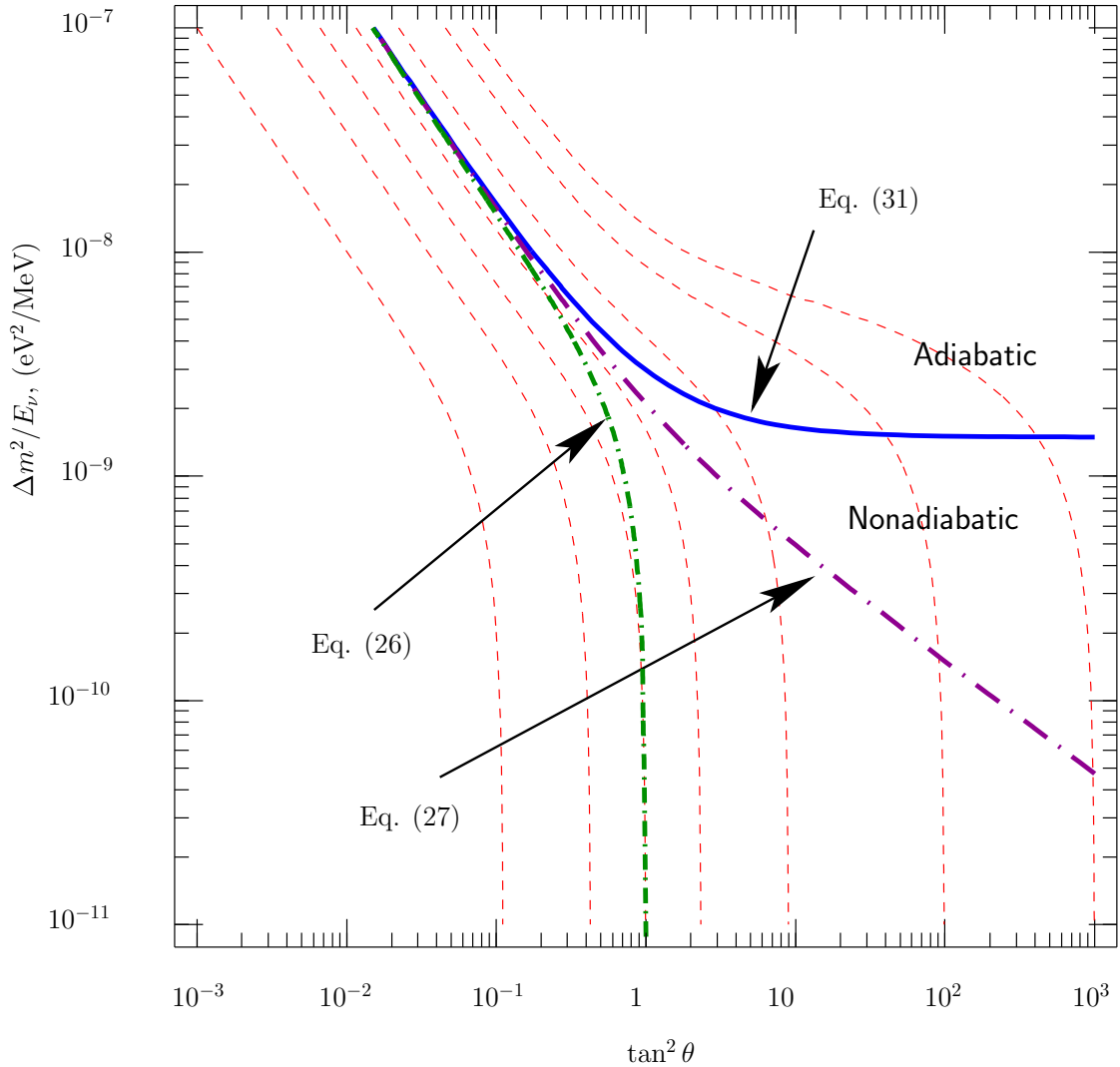}
    \caption{The boundary between the adiabatic and nonadiabatic
      regions in the case of the exponential profile, as predicted by
      Eq. (\ref{eq:oldad0}), Eq. (\ref{eq:oldad}), and
      Eq. (\ref{eq:adiabaticity}). The dashed curves
      show the contours of constant $P_c$. For large angles
      Eq. (\ref{eq:adiabaticity}) provides a better description of
      the boundary.}
    \label{fig:adiabaticity}
  \end{center}
\end{figure}

\subsection{Adiabaticity criterion for large mixing angles}

At last, we shall formulate the adiabaticity criterion that is valid
for all values of the mixing angle. From the preceding discussion it
should be obvious that the criterion of Eq.~(\ref{eq:simple_res2}),
\begin{equation}
  \label{eq:oldad0}
  |\Delta_m/\dot\theta_m|_{\theta_m=\pi/4}\gg 1,
\end{equation}
should only work in the limit $\theta\ll 1$. At the very least, the
quantity $\Delta_m/\dot\theta_m$ should be evaluated at the true point
of the maximal violation of adiabaticity. As an estimate, we can
approximate it by the point halfway between $\theta$ and $\pi/2$ (see
Eq.~(\ref{eq:maxangle})). We then get
\begin{equation}
  \label{eq:oldad}
  |\Delta_m/\dot\theta_m|_{\theta_m\sim\pi/4+\theta/2}\gg 1.
\end{equation}

This is better, but still not quite right for large $\theta$. 
The key is to express the information contained in the system of two
evolution equations in a single equation. Eq.~(\ref{eq:trivH})
contains redundant information: while it contains four real functions,
the actual number of degrees of freedom is only two, since the
normalization of the state is fixed and the overall phase is
physically irrelevant. It will be easier to formulate the criterion in
question once the two physical degrees of freedom are isolated.

With this in mind, let us write down the evolution equation for the
ratio $s\equiv\psi_1/\psi_2$. From Eq.~(\ref{eq:massbasis}) we obtain
the following \emph{first order} equation:
\begin{equation}
  \label{eq:firstorderx}
  \frac{d s}{d x} = 2 i \Delta_m s - \dot\theta_m  (s^2+1),
\end{equation}
or
\begin{equation}
  \label{eq:firstorder}
  \frac{d s}{d \theta_m} = 2 i \frac{\Delta_m}{\dot\theta_m } s - (s^2+1).
\end{equation}
It is easy to see that the
adiabatic limit corresponds to neglecting the terms in parentheses,
while the extreme nonadiabatic limit is obtained if one neglects the
first term on the right. In the second case the solution is
$s=\cot(\theta_m)$.  Thus, the self consistent condition to have the
extreme nonadiabatic solution is $2|\Delta_m/\dot\theta_m
s|\ll(s^2+1)$, with $s=\cot(\theta_m)$,
$\theta_m=\pi/4+\theta/2$. In the opposite limit, the evolution is
adiabatic. The adiabaticity condition is then
\begin{eqnarray}
  \label{eq:adiabaticity0}
  |\Delta_m/\dot\theta_m|_{\theta_m\sim\pi/4+\theta/2} \gg 
(\tan(\pi/4+\theta/2)+\cot(\pi/4+\theta/2))/2,
\end{eqnarray}
or, upon simplification,
\begin{eqnarray}
  \label{eq:adiabaticity}
  \cos\theta |\Delta_m/\dot\theta_m|_{\theta_m\sim\pi/4+\theta/2} \gg 1.
\end{eqnarray}
Eq.~(\ref{eq:adiabaticity}) is valid for all values of the
mixing angle.

To illustrate this result, let us again consider the case of the
infinite exponential profile. Eq.~(\ref{eq:adiabaticity}) in this case
becomes
\begin{eqnarray}
  \label{eq:nonad_simple}
  8\Delta r_0 \sin^2\theta \gg 1.
\end{eqnarray}
The solid line in Fig. \ref{fig:adiabaticity} shows the contour of $8\Delta r_0
\sin^2\theta=1$ computed for $r_0=R_\odot/10.54$ (the fitted value in
the BP2000 solar model, see Fig. \ref{fig:profile}). For
comparison, the dash-dotted curves shows the corresponding prediction
of Eqs. (\ref{eq:oldad0}) and (\ref{eq:oldad}). The dashed curves are
the contours of constant $P_c$ computed using Eq. (\ref{eq:expPc}). It
is clear from the Figure that the description of
Eq. (\ref{eq:adiabaticity}) is correct not only for small $\theta$,
but also for $\theta\gtrsim \pi/4$.

\section{Conclusions}
\label{sec:concl}

We have seen that the neutrino evolution in the Sun can be
described as the resonance phenomenon only in the limit
$\theta\ll1$. For large values of $\theta$, including $\theta>\pi/4$,
the resonance condition does not correspond to anything physical and
one instead should simply be thinking about either adiabatic or
nonadiabatic evolution. 

Since the physics is continuous across $\theta=\pi/4$,
there is no reason to truncate the solar neutrino
parameter space at maximal mixing. In fact, the latest solar neutrino
fits correctly present their results in the $0\leq\theta\leq\pi/2$
parameter space \cite{BKS2001,Gonzalez-Garcia2001}. 

Notice, that so long as the neutrino survival probability is computed
numerically using the full solar density profile, the results are
unaffected by the considerations presented here. The value of the
preceding arguments is that they give us a correct physical picture of
what is going on when neutrinos travel through the Sun and, for
example, explain why in the case of the LMA solution the results are
not sensitive to the details of the profile in the convective zone, or
why the quasivacuum solutions smoothly extend into the region
$\theta>\pi/4$. To paraphrase Richard Feynman, we understand the
neutrino evolution only if we can predict what should happen in various
 circumstances without actually having to solve the evolution equations
each time.

Finally, it is worth mentioning an important article by A.~Messiah
\cite{Messiah1986} which was brought to the attention of the author
after Ref.~\cite{jumping} had been submitted to Physical Review. In
the article, Messiah considers the mixing angles continuously from $0$
to $\pi/2$ and determines the location of the point of maximal
violation of adiabaticity for all values of $\theta$. Even though his
adiabaticity criterion disagrees with the one presented here, it was
clearly a very important work that, unfortunately, was almost
completely forgotten in the subsequent literature.


\section*{Acknowledgments}

I would like to thank the organizers for putting together a very
stimulating workshop. I have benefitted greatly from many enlightening
conversations with Plamen Krastev. I am also indebted to John Bahcall
for his support and valuable comments on the draft.  This work was
supported by the W.M.~Keck Foundation.

\end{document}